# Compressive multi-beam scanning transmission electron microscopy

*Akira Yasuhara[1], Takumi Sannomiya[2*], Ryoichi Horisaki[3]*

[1] JEOL Ltd., 3-1-2 Musashino, Akishima, Tokyo, 196-8558, Japan.

[2] Department of Materials Science and Engineering, School of Materials and Chemical Technologies, Institute of Science Tokyo, 4259 Nagatsuta, Midoriku, Yokohama, 226-8503, Japan.

[3] Graduate School of Information Science and Technology, The University of Tokyo, 7-3-1 Hongo, Bunkyo, Tokyo 113-8656, Japan.

**Corresponding Authors**

* Takumi Sannomiya (Email: sannomiya.t.aa@m.titech.ac.jp)

**ABSTRACT**

We demonstrate a multi-beam scanning transmission electron microscopy (STEM) imaging that integrates down-sampling with super-resolution image reconstruction via a compressive sensing framework. A custom condenser aperture with six randomly positioned circular holes is employed to produce a multi-beam STEM probe, with the beam shape and distribution tuned through defocus. While the raw multi-beam images exhibit overlapping patterns, reconstruction using Adam optimization and total variation normalization yields high-fidelity images that closely reproduce the original sample structures, even from substantially down-sampled data. The proposed approach offers a pathway toward significant acceleration of such techniques through multibeam sparse sampling and computational reconstruction potentially useful for the analytical scanning methods in general.

## INTRODUCTION

Scanning transmission electron microscopy (STEM) has become an indispensable tool in materials science, enabling direct visualization of structural and chemical information at sub-nanometer and even atomic resolutions. [1–3] By focusing a finely converged electron probe onto the sample and detecting transmitted electrons, STEM visualizes atomic arrangements, defects, and interface structures. Beyond mere structural imaging, STEM can be combined with a variety of analytical techniques, such as energy-dispersive X-ray spectroscopy (EDS), electron energy-loss spectroscopy (EELS) or cathodoluminescence (CL), to provide local information about elemental composition, electronic structure, chemical bonding, etc. [2,4,5] These capabilities have contributed to progress in diverse fields, including semiconductor devices, energy materials, catalysts, and quantum materials. Despite these advances, a persistent challenge in STEM is the relatively slow image acquisition speed, which is fundamentally constrained by the nature of point-by-point scanning processes. High-resolution STEM imaging typically demands fine pixel sampling and long dwell times per point to ensure sufficient signal-to-noise ratio, especially for weak signals in analytical methods, such as EDS or EELS. This limitation becomes particularly problematic in dose-sensitive materials, time-resolved studies, or high-throughput analysis, where minimizing electron dose and acquisition time are crucial. To address these challenges, various strategies have been explored; One prominent approach is ptychography, which utilizes a spatially extended coherent probe scanned over the sample with coarse spatial sampling, while capturing the full diffraction pattern at each position. [6] This method allows for computational reconstruction of high-resolution phase images and has demonstrated impressive results for structural imaging. Structured beam illumination has also been applied to image wider spatial frequency ranges. [7] However, ptychography requires high-speed, high-dynamic-range pixelated detectors and generates large volumes data sets. Moreover, it is inherently limited to coherent electron detection and cannot be applied to analytical signals such as photons or inelastically scattered electrons, precluding its use in EDS, EELS or CL.

In this study, we propose a novel imaging strategy that utilizes a multi-beam scanning probe in combination with bucket detection, enabling the reconstruction of high-pixel-resolution images from sparsely sampled data using a framework of compressive sensing. (Figure 1) A multi-hole aperture is inserted in the condenser lens system of a STEM instrument to generate multiple probe beams under a controlled defocus condition, resulting in structured multi-probe illumination over the sample. By leveraging prior knowledge of the beam profile, the acquired signals can be computationally deconvoluted using sparsity-based reconstruction algorithms. We note that modifications of the beam profile, or beam shaping approaches, have been employed to generate vortex beams or to enhance contrast (e.g., using phase plates), [8–13] but have not yet been applied to such multiplexing. Importantly, since this proposed method is based on bucket detection, it can be extended beyond structural imaging to include analytical techniques, offering a promising route for accelerating beam-scanning methods, even including scanning transmission electron microscopy (SEM).

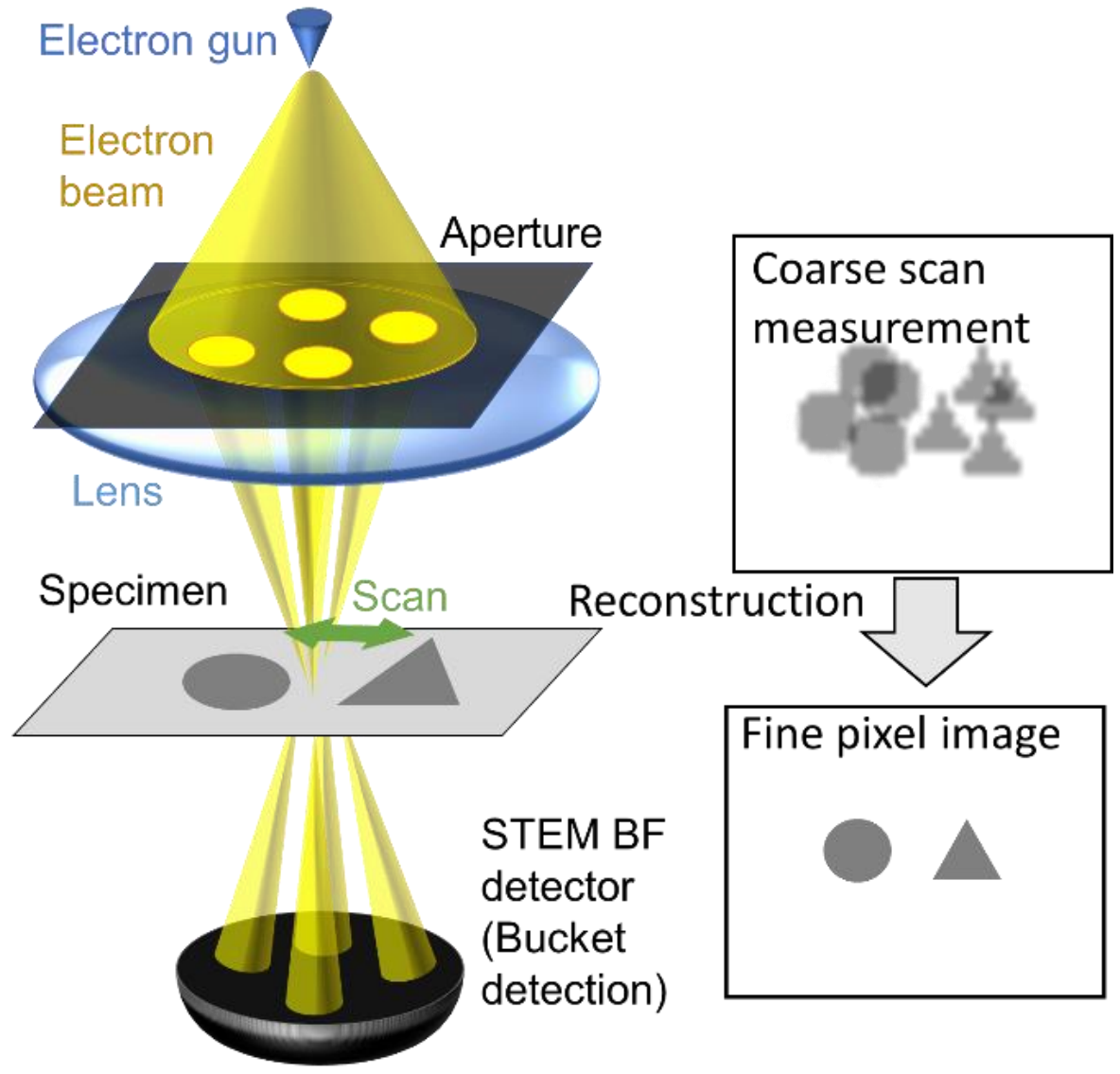

**Figure 1.** Concept of compressive multibeam scanning transmission electron microscopy. By using a multibeam probe, a high-pixel-resolution image can be reconstructed from a coarsely sampled scan through a compressive sensing approach.

## METHOD

### *Instrumentation*

A multi-hole aperture was fabricated by focused ion beam (FIB) milling on a free-standing 10 μm-thick gold (Au) film. [12] This aperture was installed in the condenser lens aperture module of a JEM-ARM200F instrument (JEOL Ltd., Japan) equipped with a symmetric objective lens pole piece (Figure 2). All the measurements were performed at the acceleration voltage of 200 kV. Six circular holes, each approximately 8 μm in diameter, were randomly distributed across a gold membrane. This random distribution is critical for sparse measurements, as the structured probe must "sense" a wide range of spatial frequencies in order to capture the complete information of the sample.

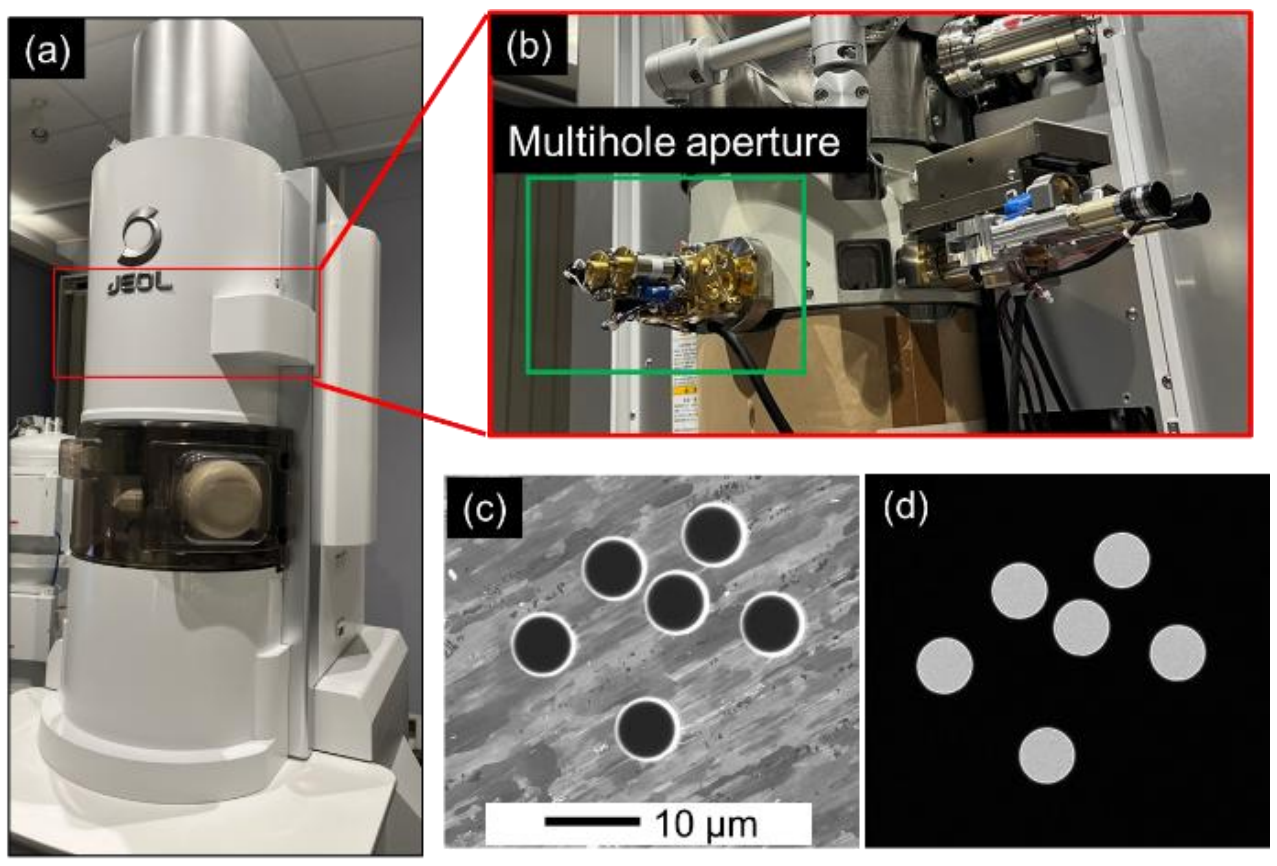

**Figure 2.** Condenser lens (CL) aperture module of the JEM-ARM200F instrument. (a, b) Photographs of the microscope and the CL aperture module. (c) Secondary ion microscopy image and (d) transmission electron microscopy (TEM) image of the fabricated multi-hole aperture.

***Data acquisition***

STEM observations were performed in bright-field (BF) mode using a conventional BF detector, which collects signals from all aperture openings (i.e., bucket detection). The probe-forming condenser lens system, including an aberration corrector with hexapoles turned off, was used to adjust the convergence angle such that the probe size was on the order of a few nanometers. This relatively low convergence angle configuration enabled direct imaging of the probe shape in the real space with minimal influence from aberrations in the imaging lens system. Real-space probe spot images were acquired in transmission electron microscopy (TEM) mode. To control the shape of the illumination probe, the objective lens defocus was varied. It should be noted that, due to the symmetric in-lens design of the objective lens pole piece, the defocus applied in the imaging path results in a spot image with twice the defocus value compared to the actual probe on the sample. The beam scan is performed by conventional scan coils.

The electron optical settings are summarized in Figure 3. The magnification of the aperture image was tuned to ensure that spherical aberrations in the illumination system is not significant (Figure 3a–c). For demonstration purposes, a gold nanoparticle sample was used, as shown in Figure 3d.

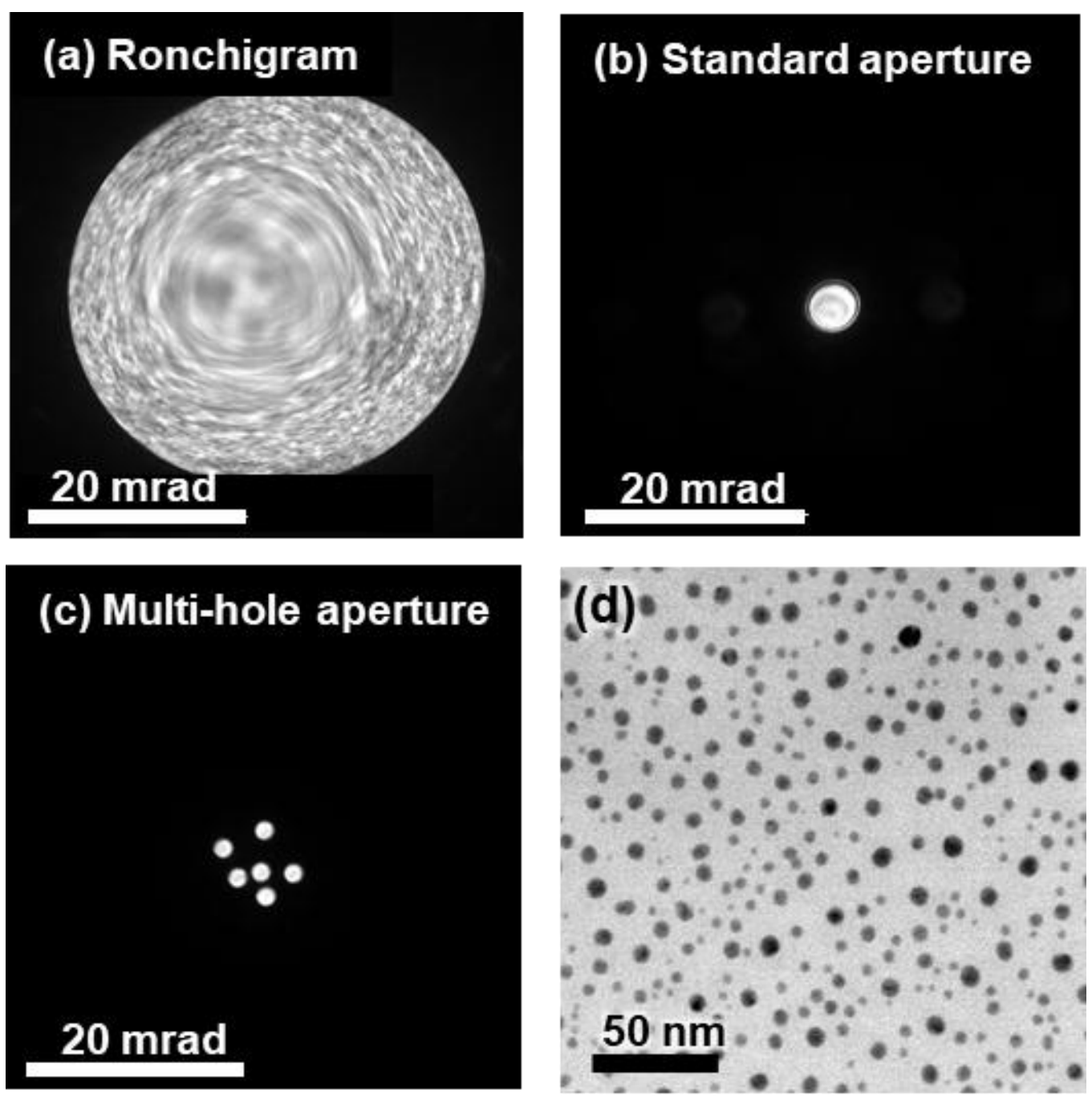

**Figure 3.** Electron optical conditions used in the experiment. (a) Ronchigram showing the magnitude of spherical aberration. (b) Corresponding single-aperture image on the detection plane, used for reference image acquisition. (c) Multi-hole aperture image on the detection plane. (d) Bright-field reference image of the gold nanoparticle sample, acquired using a single hole aperture shown in (b).

### *Image Reconstruction*

In the proposed multi-beam scanning method, the object *o* is convolved with the illumination probe *p*. Here we consider both *o* and *p* are real functions for the intensity-based measurement. The resulting convolution is then down-sampled due to the coarse scanning of the probe. This measurement process can be expressed as:

$$i = D[o * p], \qquad (1)$$

where *i* denotes the captured image, * represents the convolution operator, and *D*[•] indicates the down-sampling operator.

The inversion of Eq. (1) is an ill-posed problem. To address this, we reconstruct the object *o* by solving the following optimization problem with sparsity regularization based on compressive sensing [14–17]:

$$\arg\min_{\hat{o}} \|i - D[\hat{o} * p]\|_2^2 + \alpha \mathrm{TV}[\hat{o}], \qquad (2)$$

where $\hat{o}$ denotes the estimated object, $\|\bullet\|_2$ represents the $\ell_2$ norm, $\alpha$ is a parameter controlling the strength of the regularization, and TV[•] denotes the total variation (TV) norm. [18] In the present reconstruction process, $\alpha$ was set to $10^{-4}$. This optimization problem is solved using gradient descent with the Adam optimizer to ensure stable and fast convergence. [19]

Before conducting the optimization described in Eq. (2), the illumination probe *p* is estimated by solving the following optimization problem:

$$\arg\min_{\hat{p}} \|i' - o' * \hat{p}\|_2^2, \qquad (3)$$

where $\hat{p}$ denotes the estimated probe, and *i'* is a non-down-sampled image captured by scanning the probe at full resolution on a calibration target *o'*. The calibration target *o'* does not need to be identical to the actual imaging object *o*; it is captured separately using a conventional single beam with the standard aperture under the full-resolution scanning condition. Gradient descent with the Adam optimizer is also employed to solve this optimization problem.

## RESULTS AND DISCUSSION

### *Probe shape evaluation*

A representative multi-beam imaging result is shown in Figure 4. The obtained STEM-BF image (Figure 4a) appears as a superposition of multiple single-probe images. This is consistent with the directly observed probe shape under the same optical conditions, which shows multiple probe spots corresponding to the aperture openings, as seen in Figure 4b. The defocus value was halved to match the image plane defocus to that at the sample plane, accounting for the symmetric in-lens objective design. (See Supplementary Material for the details) To evaluate the probe shape more clearly, we performed a simple single-step deconvolution of the multi-beam image (Figure 4a) using the reference image acquired with a single aperture (Figure 3d). As deconvolution tends to amplify background noise, Wiener filtering was applied to suppress it. The resulting deconvoluted image (Figure 4c) exhibits a six-dot pattern that roughly corresponds to the intensity distribution observed in the direct spot image of Figure 4b. While the deconvoluted probe shape is discernible, the image quality remains insufficient for use in high-fidelity image reconstruction.

In contrast, the integrative probe estimation approach of Eq. 3 yields a smoother background and a clearer probe shape, as shown in Figure 4d. The resulting point spread function (PSF) pattern differs slightly from the directly observed camera image, likely due to inaccuracies of the camera image in magnification, rotation, or residual aberrations such as defocus shift, astigmatism, and spherical aberration. The variation of the intensity as well as the size of the reconstructed six beam spots are due to the limited signal to noise in the images. Since the estimated PSF is derived from two real-space images without requiring any post-adjustments (e.g., scaling or rotation), it is considered more reliable than the camera-based image. Therefore, in the following sections, we use only the estimated PSF obtained from the reference image under each imaging condition.

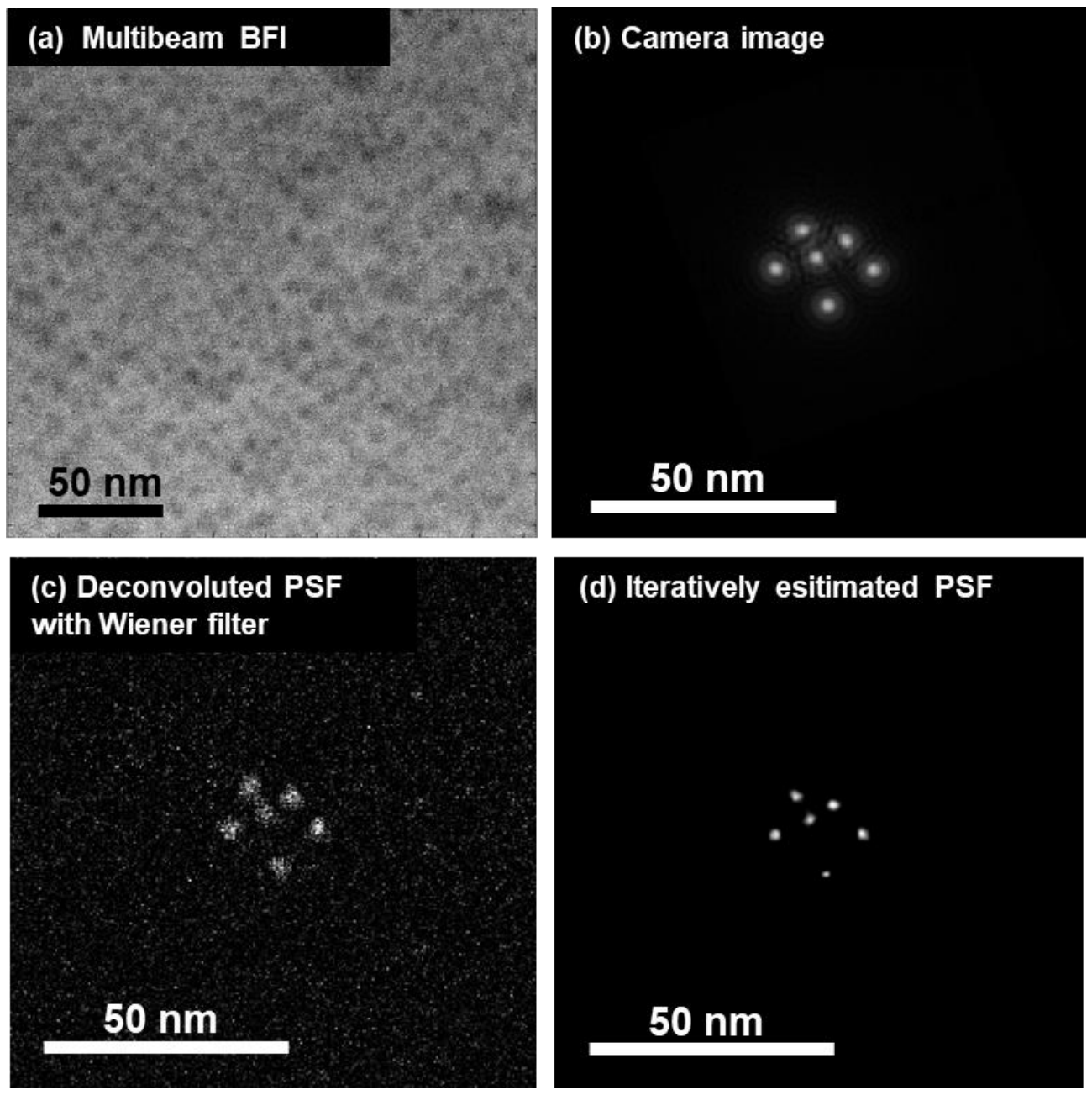

**Figure 4.** Probe shape evaluation under a representative defocus condition. (a) Multi-beam BF STEM image. (b) TEM-mode camera image of the probe, acquired with half the defocus to correspond to the probe defocus at the sample plane. (c) PSF obtained by deconvolution with Wiener filtering, using the reference image from Figure 3d. (d) PSF estimated by gradient descent with the Aam optimizer.

### *Image reconstruction from down-sampled data*

Using the estimated PSF of Figure 4, we reconstructed the sample information from the multibeam images to evaluate image resolution enhancement from down-sampled data. A $512 \times 512$ pixel image was reconstructed from measurement images at four different sampling densities, namley $512 \times 512$, $256 \times 256$, $170 \times 170$, and $128 \times 128$ pixels. The corresponding results are shown in Figure 5. The results show a clear trade-off between sampling density and image quality, which is a fundamental challenge. While reducing the number of scanned pixels accelerates data acquisition and minimizes electron dose, a critical consideration for beam-sensitive samples, it typically results in a loss of image resolution and increased noise. Our approach, however, enables the image reconstruction preserving key structural features such as ~5 nm particles, even from significantly down-sampled data (as low as $170 \times 170$ pixels, or approximately 1/9 of the original). This robustness indicates that essential spatial information can be retained and reconstructed despite under-sampling.

The noisy features observed in the reconstructed images arise from a reduced signal-to-noise ratio, caused by the smaller effective aperture area and the corresponding lower beam current. (See Supplementary Material Fig. S1 for the noise effect evaluation.) The noise amplification becomes more pronounced with increased down-sampling, despite the application of TV regularization. While TV normalization effectively suppresses background noise, it can also introduce overly harsh contrast, potentially obscuring subtle image details. Additionally, excessive down-sampling relative to the number of probe spots in the multibeam pattern can significantly degrade image quality, as some sample regions may not be illuminated at all. Increasing the number of beams (i.e., using more holes in the aperture) would help mitigate this issue by providing more comprehensive spatial coverage.

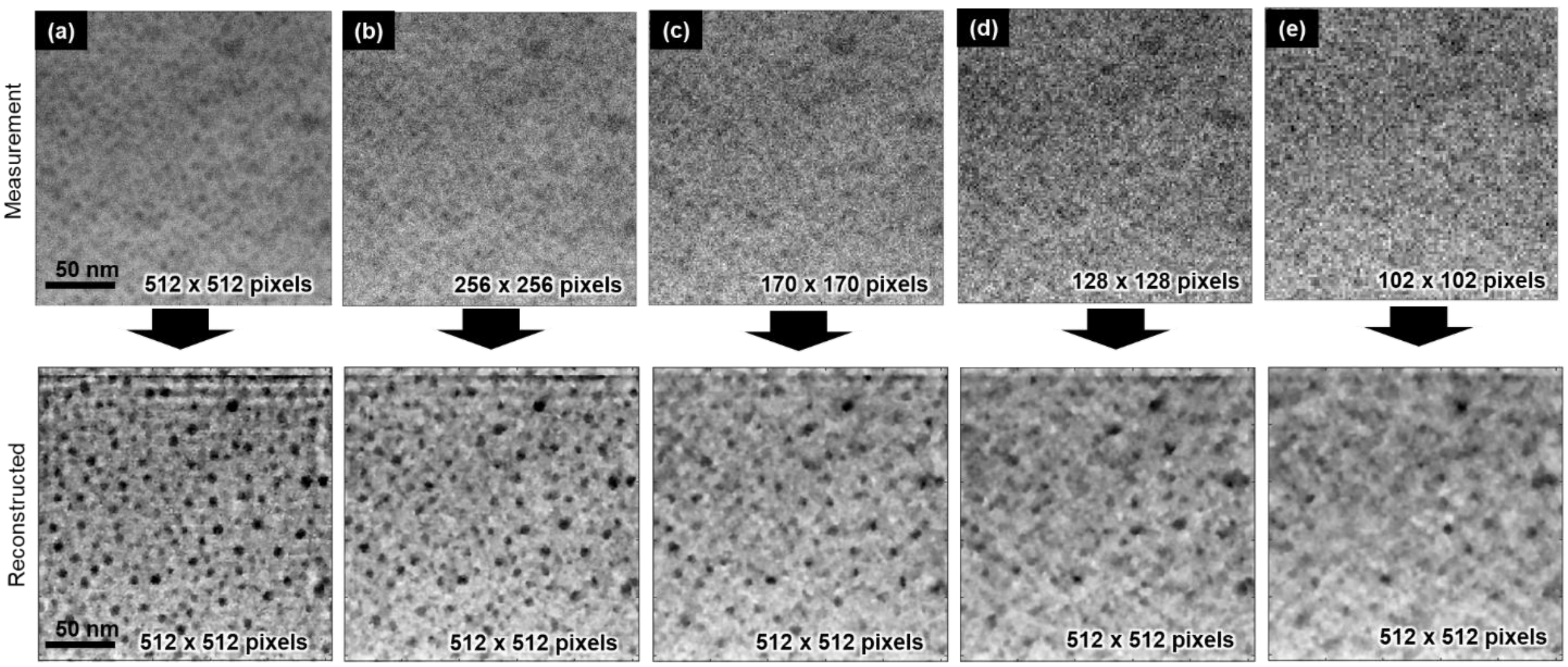

**Figure 5.** Reconstruction of 512 × 512 pixel images from datasets with varying sampling densities: (a) 512 × 512, (b) 256 × 256, (c) 170 × 170, (d) 128 × 128 pixels, and (e) 102 × 102 pixels. The images are acquired under the probe condition of Figure 4.

To evaluate the similarity to the original reference image (Fig. 3d), we applied the structural similarity index measure (SSIM) to the reconstructed images of Fig.5. [20] Since SSIM evaluation is strongly influenced by the salt-and-pepper type noises, which inevitably exist in our experimental images especially in the background contrast, we applied most weight on the structure component than luminance or contrast component and set the evaluation window size to 6 pixels. The edges of the image frame influenced by the Fourier transformation processes were excluded. The obtained SSIM values are 0.531, 0527, 0.474, 0.461, and 0.430, respectively from the higher density sampling data to the lower. The 256 pix data gives a similar SSIM value to that of the full sampling, ensuring the validity of this down sampling approach. However, excessively reduced sampling gives low values, which is apparent from the look of the image. The corresponding local SSIM maps (See Supplementary Material, Fig.S2) shows close to 1 values (good match) around the particles but lower values mostly at the flat background areas where no particles exist. This indicates that the tiny random contrasts in the support film, which are not reconstructed well, reduces the total SSIM values.

***Probe shape dependence***

Since down-sampled image acquisition at 1/4 of the original density (256 × 256 pixels) reproduces the image relatively well, as described in the previous section, we used this sampling rate to compare different illumination conditions. Figure 6 presents the reconstruction results under five different defocus settings. At larger defocus values (Figure 6a), the individual beams in the multibeam probe are more widely separated, causing greater overlap in the resulting image and making it more challenging to visually discern the original sample features. In contrast, smaller defocus conditions (Figure 6e) produce a more tightly focused probe, with beams closer together, resulting in "blurred" images rather than "overlapped" that are easier to interpret by eye.

Interestingly, despite the more visually overlapping images, the 512 × 512 pixel reconstructions are of higher quality at larger defocus (Figure 6a or b), compared to the ones at smaller defocus giving more discernable images by eye (Figure 6a or b). A larger defocus produces a PSF with more widely spaced intensity spots, enabling better coverage of the sample's spatial frequencies than probes with smaller defocuses. In other words, the spatial frequencies necessary to accurately reconstruct fine details are more fully sampled when the probe beams are sufficiently separated. Indeed, the comparison of the beam shape and the autocorrelation image of the reference image in Fig S3 in SM shows that the probe with the separate beams effectively overlaps the lower frequency features (away from the center in autocorrelation) of this sample. Overall, these results highlight the importance of optimizing defocus and probe configuration in multibeam STEM imaging.

We evaluated SSIM also for this result. (Local SSIM maps are given in the Supplementary Material) The total SSIM values for the Fig. 6 data are 0.543, 0.493, 0.490, 0.485, and 0.523 respectively for the panels a-e. The SSIMS values also tend to be better for the more separated beams in PSF. The large SSIM value for the most focused data (panel e) seems due to the lower background contrast in the support film part rather than the particle positions, according to the local SSIM map (Fig. S4 in SM). This is again related to the match of the probe shape and the sample feature; The focused probe effectively captures the high frequency components while moderately defocused beams tend to probe featureless frequencies (Fig. S2 in SM).

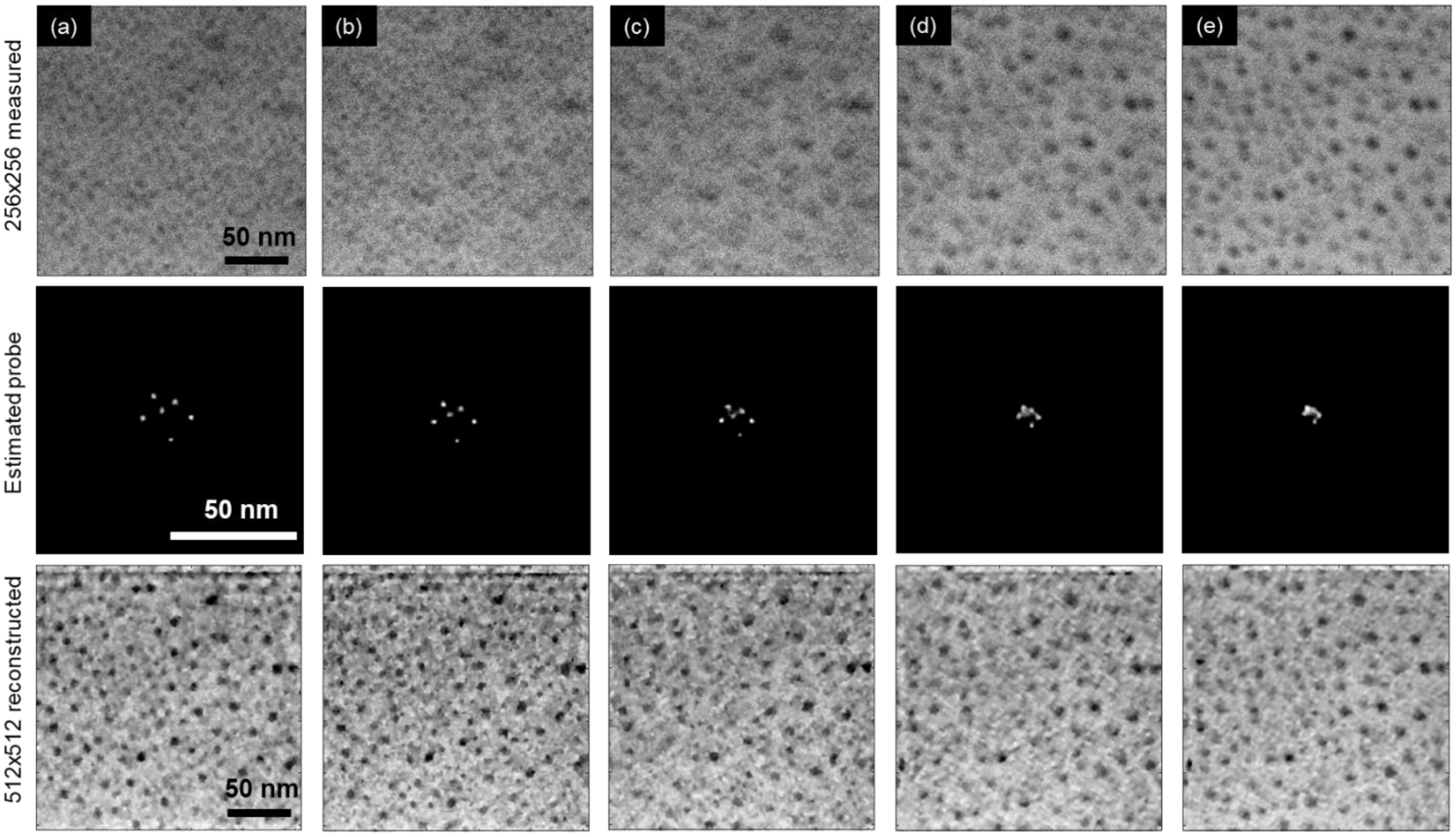

**Figure 6.** Probe shape dependence of the image reconstruction. Defocus decreases from (a) to (e), as indicated by the more overlapped images in the upper row and the more widely separated intensity spots in the PSF shown in the middle row for panel (a). The reconstructed images are displayed in the bottom row.

## CONCLUSIONS

We have demonstrated multi-beam STEM bright-field imaging combined with down-sampling and super-resolution image reconstruction, enabled by a compressive sensing approach. A condenser aperture containing randomly distributed circular holes was used to generate a multibeam STEM probe, with the beam shape and distribution controlled via defocus. Although the experimentally obtained multibeam images appear as overlapping superpositions, the reconstruction algorithm based on Adam optimization and TV normalization successfully recovered high-quality sample images that reasonably represent the original structures, even from significantly down-sampled datasets. Our results indicate that probe shapes featuring more widely separated beam spots provide better image reconstruction. Thus, balancing beam separation to enhance frequency sampling seems crucial for achieving high-quality reconstructions from sparse data. Future work could explore adaptive defocus control or dynamic aperture designs to optimize multibeam configurations and further improve imaging performance. Although we demonstrated rather low-magnification imaging here, higher magnification imaging should be possible when the beam shape is properly designed. Since the present image reconstruction is based on the real amplitude (intensity) imaging, further development would be needed when phase contrasts are involved.

Importantly, since this approach relies on bucket detection of intensities rather than pixelated detectors like ptychography, it is directly applicable to analytical STEM and SEM methods such as EDS, EELS, and CL methods, all of which typically require long acquisition times. The presented result suggests the potential for significant acceleration of these analytical techniques through multibeam sparse sampling and computational reconstruction.

## ACKNOWLEDGMENT

This work is financially supported by JSPS Kakenhi (JP23K26567, JP23H05444, 24H00400, 25K22227), JST FOREST (JPMJFR213J, JPMJFR2448), JST CREST (JPMJCR25I3), Asahi Glass Foundation, and SECOM Science and Technology Foundation.

## Author Contributions

The manuscript was written through contributions of all authors. All authors have given approval to the final version of the manuscript.

## REFERENCES

[1] O. L. Krivanek, M. F. Chisholm, M. F. Murfitt, and N. Dellby, Scanning transmission electron microscopy: Albert Crewe's vision and beyond, Ultramicroscopy **123**, 90 (2012).

[2] C. Ophus, Quantitative Scanning Transmission Electron Microscopy for Materials Science: Imaging, Diffraction, Spectroscopy, and Tomography, Annu. Rev. Mater. Res. **53**, 105 (2023).

[3] R. Ishikawa, T. Tanigaki, and Y. Fukuda, Resolution does matter, Microscopy **72**, 65 (2023).

[4] A. Maigne and R. D. Twesten, Review of recent advances in spectrum imaging and its extension to reciprocal space, J. Electron Microsc. (Tokyo) **58**, 99 (2009).

[5] A. Yasuhara, M. Shibata, W. Yamamoto, I. Machfuudzoh, S. Yanagimoto, and T. Sannomiya, Momentum-resolved EELS and CL study on 1D-plasmonic crystal prepared by FIB method, Microscopy **73**, 473 (2024).

[6] C. Huang, J. S. Kim, and A. I. Kirkland, Cryo-electron ptychography: Applications and potential in biological characterisation, Curr. Opin. Struct. Biol. **83**, 102730 (2023).

[7] H. Tamaki and K. Saitoh, Near-field electron ptychography using full-field structured illumination, Microscopy **74**, 10 (2025).

[8] R. Danev and K. Nagayama, Transmission electron microscopy with Zernike phase plate, Ultramicroscopy **88**, 243 (2001).
[9] J. Haas, N. Rieger, M. Schlegel, K. Strobel, and J. C. Meyer, Nanometer-scale electron beam shaping with thickness controlled and stacked nanostructured graphite, Appl. Phys. Lett. **124**, 233501 (2024).
[10] E. Rotunno, A. H. Tavabi, E. Yucelen, S. Frabboni, R. E. Dunin Borkowski, E. Karimi, B. J. McMorran, and V. Grillo, Electron-Beam Shaping in the Transmission Electron Microscope: Control of Electron-Beam Propagation Along Atomic Columns, Phys. Rev. Appl. **11**, 044072 (2019).
[11] S. M. Lloyd, M. Babiker, G. Thirunavukkarasu, and J. Yuan, Electron vortices: Beams with orbital angular momentum, Rev. Mod. Phys. **89**, 035004 (2017).
[12] A. Yasuhara, F. Hosokawa, S. Asaoka, T. Akiyama, T. Iyoda, C. Nakayama, and T. Sannomiya, Semicircular-aperture illumination scanning transmission electron microscopy, Ultramicroscopy **270**, 114103 (2025).
[13] M. Malac, S. Hettler, M. Hayashida, E. Kano, R. F. Egerton, and M. Beleggia, Phase plates in the transmission electron microscope: operating principles and applications, Microscopy **70**, 75 (2021).
[14] E. J. Candes and M. B. Wakin, An Introduction To Compressive Sampling, IEEE Signal Process. Mag. **25**, 21 (2008).
[15] R. Baraniuk, Compressive Sensing [Lecture Notes], IEEE Signal Process. Mag. **24**, 118 (2007).
[16] J. Romberg, Compressive Sensing by Random Convolution, SIAM J. Imaging Sci. **2**, 1098 (2009).
[17] R. Horisaki and J. Tanida, Multi-channel data acquisition using multiplexed imaging with spatial encoding, Opt. Express **18**, 23041 (2010).
[18] L. I. Rudin, S. Osher, and E. Fatemi, Nonlinear total variation based noise removal algorithms, Phys. Nonlinear Phenom. **60**, 259 (1992).
[19] D. P. Kingma and J. Ba, *Adam: A Method for Stochastic Optimization*.
[20] Z. Wang, A. C. Bovik, H. R. Sheikh, and E. P. Simoncelli, Image quality assessment: from error visibility to structural similarity, IEEE Trans. Image Process. **13**, 600 (2004).

# Supplementary Material for

# Compressive multi-beam scanning transmission electron microscopy

*Akira Yasuhara*[1], *Takumi Sannomiya*[2*], *Ryoichi Horisaki*[3]

[1] JEOL Ltd., 3-1-2 Musashino, Akishima, Tokyo, 196-8558, Japan.

[2] Department of Materials Science and Engineering, School of Materials and Chemical Technologies, Institute of Science Tokyo, 4259 Nagatsuta, Midoriku, Yokohama, 226-8503, Japan.

[3] Graduate School of Information Science and Technology, The University of Tokyo, 7-3-1 Hongo, Bunkyo, Tokyo 113-8656, Japan.

**Corresponding Authors**

* Takumi Sannomiya (Email: sannomiya.t.aa@m.titech.ac.jp)

## Defocus in camera images

Because the objective lens of STEM used in this study is a symmetric in-lens type (sample is situated in the middle of the magnetic field), changing the defocus shifts the focus of both the for-field and post-field of the objective lens. Therefore, the focus change on the sample is doubled in the transmitted image captured by the camera since the transmitted image experiences the defocus of the post-field. Therefore, to properly correspond the defocus on the specimen to that on the camera, the value should be halved.

## Evaluation of noise effect by simulation

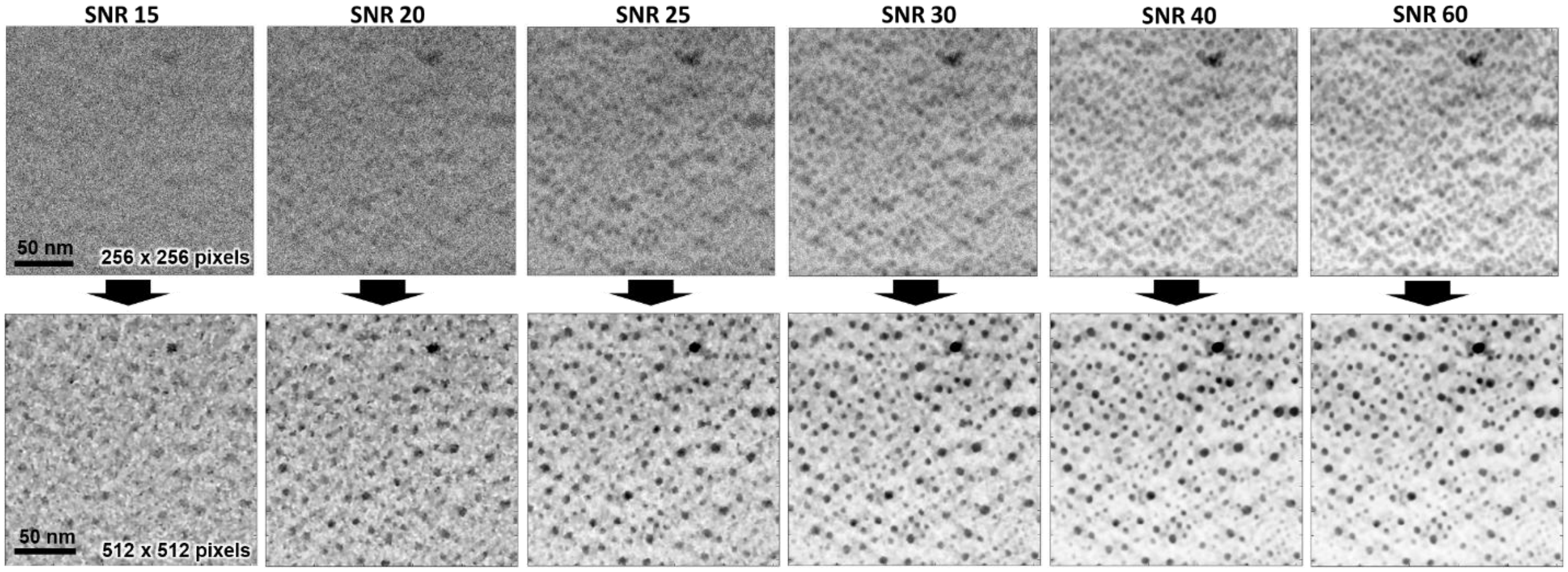

**Figure S1.** Robustness to the noise in image reproduction. The reference STEM image in Fig. 3d was convoluted with the probe shown in Fig. 4d to generate a simulated multi-probe image, to which additive white Gaussian noise was applied. The signal-to-noise ratio (SNR), which quantifies the noise level, is defined as SNR = 10 $\log_{10}$ ($P_{\text{signal}}$ / $P_{\text{noise}}$), where $P$ denotes the average power of the signal and noise, respectively. The top row shows simulated images down-sampled by a factor of two with increasing noise levels, and the bottom row shows the corresponding

reconstructed images. The quality of the experimental image in Fig. 4a is approximately equivalent to an SNR of 20–25.

### SSIM evaluation results for down-sampled data

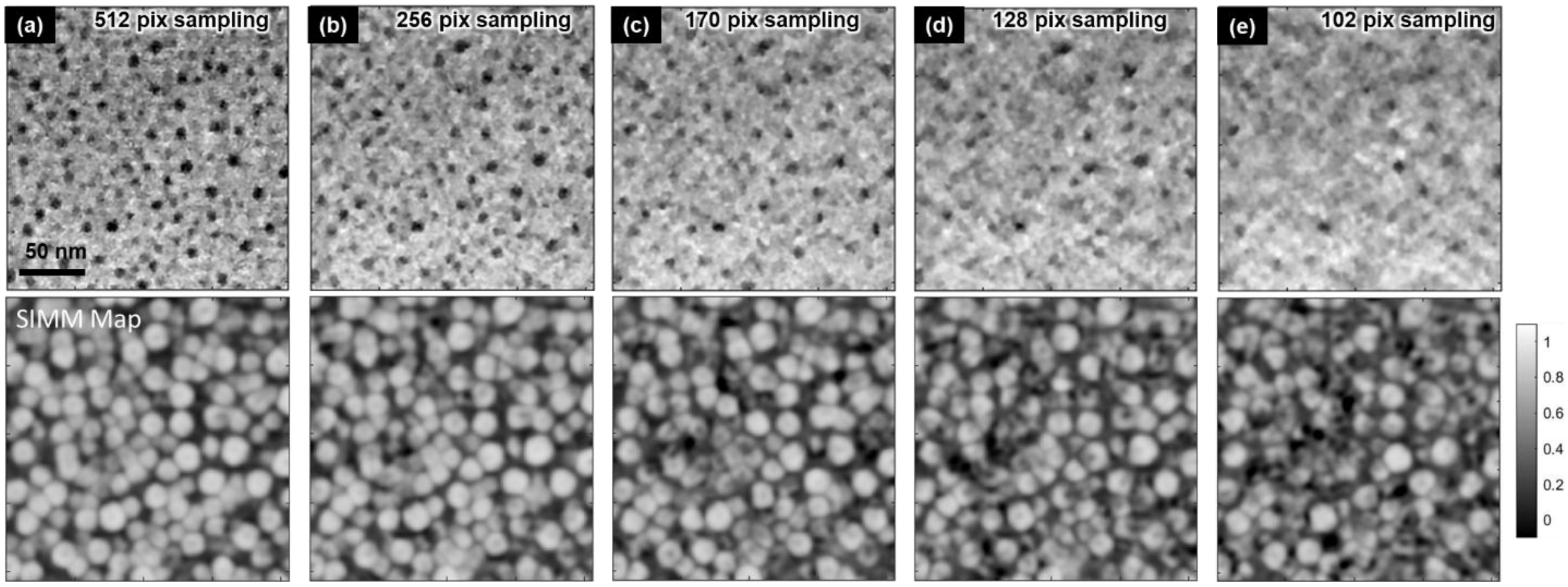

**Figure S2.** SSIM evaluation results of the data of the down-sampled data in Fig. 5 in the main text. The cropped images to exclude the artifact fringes are shown on the top and the local SSIM maps are shown on the bottom. The total SSIM values are 0.531, 0527, 0.474, 0.461, and 0.430, respectively for panels a-e.

## Comparison of probe shape and sample spatial frequency features

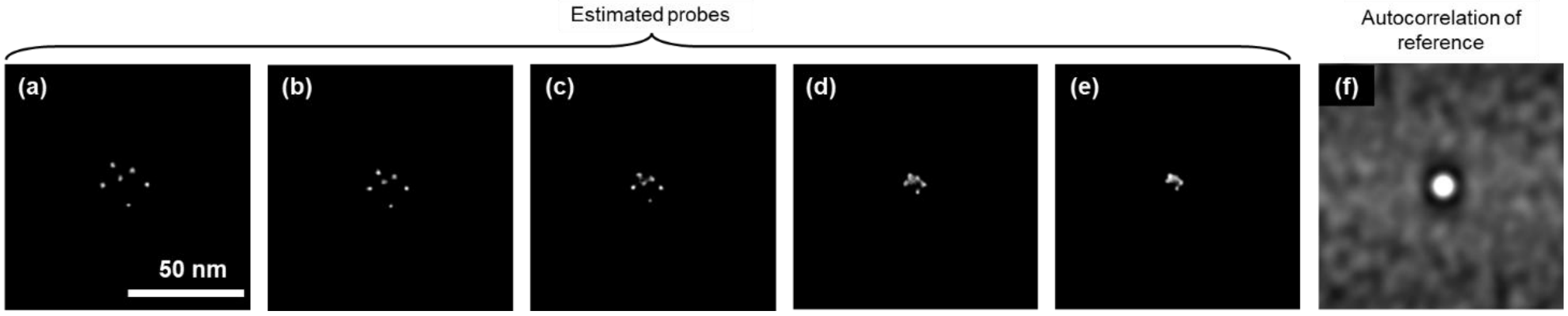

**Figure S3.** Comparison of (a-e) the probe shapes of Fig.6 and (f) the autocorrelation of the reference BF image obtained by the single aperture with a full scan. The same scale applies to all panels.

## SSIM evaluation results for different probes

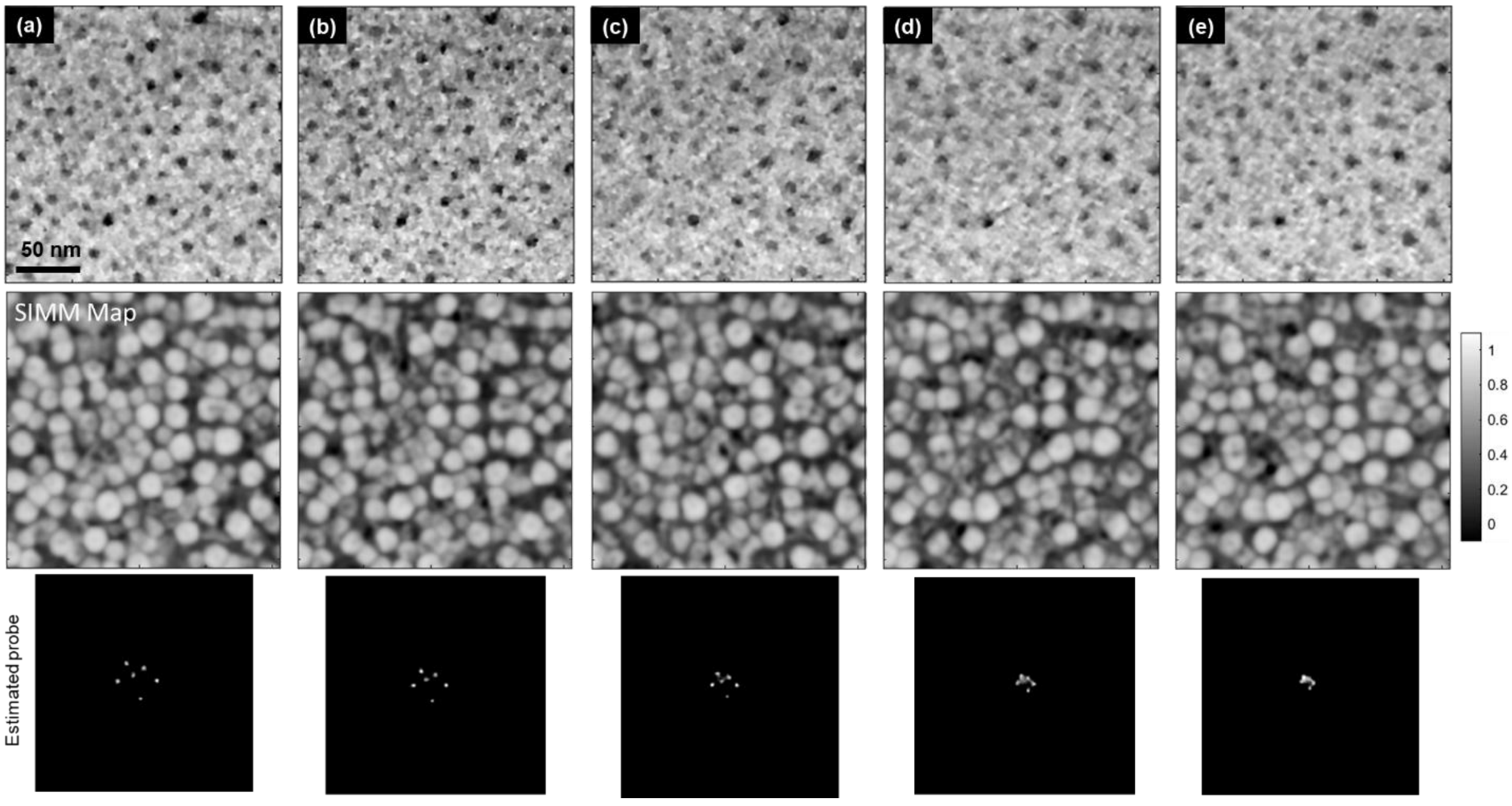

**Figure S4.** SSIM evaluation results of the data of the different probe (defocus) data in Fig. 6 in the main text. The cropped images to exclude the artifact fringes are shown in the top row and the local SSIM maps are shown in the middle row. In the bottom row the probe images are shown for comparison. The total SSIM values are 0.543, 0.493, 0.490, 0.485, and 0.523 respectively for the panels a-e.